\documentclass[aps,prl,twocolumn,superscriptaddress,floatfix,notitlepage]{revtex4-2}
\usepackage[utf8]{inputenc}
\usepackage{amsmath,amsfonts,amssymb,amsthm,bm,color,dsfont,graphicx,psfrag,mathtools,cancel,physics}
\usepackage{hyperref}
\usepackage{tikz}
\usetikzlibrary{positioning}
\usetikzlibrary{decorations.markings,arrows.meta}
\usepackage{pgfplots}
\pgfplotsset{compat=1.17}
\usepackage{xcolor}
\usepackage{soul}

\usepackage[export]{adjustbox}
\usepackage{natbib}
\usepackage[hybrid]{markdown}
\usepackage{mathrsfs}
\usepackage{blkarray}
\usepackage{multirow}
\usepackage[caption = false]{subfig}

\usepackage{longtable}
\usepackage{tabularx}
\usepackage{colortbl}
\usepackage{multirow}
\newcommand{\I}{\mathds{1}}

\newcommand{\QH}{\mathrm{QH}}
\newcommand{\QE}{\mathrm{QE}}

\newcolumntype{M}[1]{>{\centering\arraybackslash}m{#1}}
\newcolumntype{N}{@{}m{0pt}@{}}
\newcolumntype{C}[1]{>{\centering\arraybackslash}p{#1}}

\def\boti  {\,{\boxtimes}\,} 
\def\Is    {\mathcal I \hspace*{-1.1pt}s}
\def\oti   {\,{\otimes}\,}
\def\Ufour {\mathcal Z_8^{(1/2)}}
\def\sqi  {\,{1/\sqrt{2}}\,} 

\begin{document}

\title{
%Microscopic derivation of fusion rules of topological defects in quantum Hall systems\\
%	Fusion of Topological Excitations in Fractional Quantum Hall Systems: \\ A~Microscopic Approach
Unveiling Topological Fusion in Quantum Hall Systems from Microscopic Principles
}

\author{Arkadiusz Bochniak}
\email{arkadiusz.bochniak@uj.edu.pl}
\affiliation{%
Institute of Theoretical Physics, Jagiellonian University, {\L}ojasiewicza 11, Krak{\'o}w, 30-348, Poland
}%
\affiliation{%
Max-Planck-Institut f{\"u}r Quantenoptik,Hans-Kopfermann-Str.~1, Garching, 85748, Germany
}%
\affiliation{%
Munich Center for Quantum Science and Technology, Schellingstraße~4, M{\"u}nchen, 80799, Germany
}%

\author{Shinsei Ryu}
\email{shinseir@princeton.edu}
\affiliation{
Department of Physics, Princeton University, Princeton, New Jersey, 08544, USA
}%

\author{J\"urgen Fuchs}
\email{juergen.fuchs@kau.se}
\affiliation{
Teoretisk Fysik, Karlstads Universitet, Karlstad, Sweden
}%

\author{Gerardo Ortiz}
\email{ortizg@iu.edu}
\affiliation{Institute for Advanced Study, Princeton, NJ 08540, USA}
\affiliation{% 
Department of Physics, Indiana University, Bloomington IN 47405, USA}%

\date{\today}% It is always \today, today,

\begin{abstract}
	Establishing the fusion rules of anyonic quasiparticles in fractional quantum Hall fluids is essential for understanding their underlying topological order. Building on the conjecture that key topological properties are encoded in the “DNA” of candidate many-body wave functions—that is, the pattern of dominant orbital occupations restricted to a finite number of lowest Landau levels—we propose a combinatorial framework that derives these fusion rules directly from microscopic data. By extending Schrieffer’s counting argument and introducing classes of topological excitations, our framework provides a unified route to the fusion rules for both Abelian and non-Abelian excitations.  This approach elucidates the emergence of topological features from first principles in both fermionic and bosonic systems. In particular, we unveil the previously unknown fusion rules of the Hafnian fluid.
\end{abstract}

\maketitle

%\tableofcontents

{\it Introduction.--} Fractional quantum Hall (FQH) fluids are among the most striking examples of quantum phases that defy description by local order parameters, being instead characterized by global topological properties of their ma\-ny-body wave functions. Their elementary excitations, known as anyons, carry fractional charge and obey nontrivial exchange statistics, which in some cases is non-Abe\-li\-an. 
A central focus of research for decades has been to uncover the fractional nature directly from many-body wave functions
\cite{Laughlin83, Anderson83, Arovas84}  
-- multivariate polynomials of electron coordinates that, when projected to the lowest Landau level (LLL), 
become holomorphic. In particular, trial or model wave functions have been employed to study key properties of FQH states, including braiding statistics, fusion rules, and fractional charge
\cite{Laughlin83,Moore91,Bernevig08}.
These explicit trial many-body states capture the essential physics, serving as blueprints that connect microscopic Hamiltonians to effective topological quantum field theory (TQFT) descriptions. 
They can be constructed, for example, using conformal field theory (CFT) 
\cite{Moore91, Hansson_2017},
composite fermion theory 
\cite{Jain_book}, 
or Jack polynomials and their variants 
\cite{Bernevig08,Bernevig08a}.

In spite of these developments, a systematic method for
deriving the topological properties directly from the microscopic
wave functions remains elusive. In the present note, we aim to
develop a microscopic derivation of the fusion rules of
topological excitations in FQH fluids.
Specifically, in our approach we focus on what we call the root pattern or ``DNA'' of FQH ground states 
\cite{Bernevig08,DNA18}. 
DNA here refers to the pattern of dominant orbital occupations in the LLL. Together with its squeezing hierarchy,
that is, the set of configurations obtained by successive squeezing operations that preserve angular momentum,
the DNA encodes the full many-body wave function. Moreover, in the Tao–Thouless (thin-cy\-lin\-der/torus) limit, 
the DNA itself is an exact ground state.
It has been conjectured that certain topological characteristics of FQH fluids can be inferred from their DNA, and this idea has been explored from various perspectives 
\cite{Ortiz13,Ahari23}. 
To the best of our knowledge, no significant deviations from this conjecture have been found. Our framework builds on this hypothesis and investigates its further implications. 

Central objects of our interest are interfaces separating
distinct ground-state configurations (DNAs), which act as hosts
for fundamental topological excitations, and to which we refer
as \emph{domain walls} or \emph{topological defects}.  In the Abelian regime, the concept of fractional charge is well captured by Schrieffer’s counting argument, 
initially introduced to explain soliton charges in polyacetylene \cite{Su71,Wilczek2002,boer_dal_ulfbeck_1986}. 
In FQH fluids, such domain walls correspond to localized quasiholes (QHs) or quasielectrons (QEs), 
whose presence disrupts the DNA configuration of the ground state. 

Our approach for extracting fusion rules from DNAs, thereby connecting microscopic orbital configurations to emergent topological properties, proceeds in two steps. First, we identify a set of combinatorial {\it flux-insertion rules} that generate domain walls of a given charge, which host fundamental topological excitations. 
Our method provides a general framework encompassing all charge sectors, and it applies to both bosonic and fermionic systems \footnote{For elementary charges, a different set of rules was proposed in Refs.~\cite{Ardonne_2008,Ardonne09} for certain bosonic FQH fluids}. These rules naturally define relations among domain walls. Second, we derive the fusion rules for the resulting classes. While our results can be compared and matched with TQFT and CFT descriptions when such formulations exist (see \cite{supp}), we emphasize that our microscopic framework is developed entirely within the DNA perspective and does not rely on TQFT or CFT.

Another important feature of our work is that we fully incorporate the electric $U(1)$ charge into our framework,
i.e., fractional charge and charge superselection sectors. While one may think that $U(1)$ symmetry acts as a mere enrichment of topological order, they are deeply intertwined in the context of FQH liquids. For example, the correspondence between Jack polynomials and CFTs arises only in the presence of $U(1)$ charge symmetry. Specifically, in categorical terms, the fusion category under consideration is not the Deligne product of the charge sector with another CFT, but rather the category obtained from this product via a gauging procedure involving a suitable (super)commutative algebra internal to the Deligne product. Further details are provided in the End Matter and the Supplemental Material, where we discuss categorical aspects of non-Abelian FQH fluids, including the fermionic Moore–Read state and the $k \,{=}\, 3$ Read–Rezayi state, with particular emphasis on incorporating the $U(1)$ charge sectors.

{\it Classes of domain walls.--} 
Within the LLL, the fluid's DNA can be characterized via the root pattern. Given wave functions for $N$ particles describing the ground state and excitations of FQH fluids, one can determine their root patterns by maximizing the quantity $\Delta_J \,{=} \sum_{i=1}^N j_i^2$, where $j_i$ denotes the angular momentum of the $i$-th particle  \cite{Seidel08}. The root state of a translationally invariant incompressible fluid consists of $m$ {\it unit cells}, each of length $M$, containing $n_{\sf p}$ particles per cell, such that $N \,{=}\, n_{\sf p} \, m$. The set of center-of-mass-degenerate states on the torus forms a moduli space $\mathcal{M}$.  For instance, for the $\nu \,{=}\, 1/3$ Laughlin ground state, we have $M \,{=}\, 3$ and the moduli space reads $\{\lfloor 1\rfloor \,{=}\, (100), \lfloor 2\rfloor \,{=}\, (010), \lfloor 3\rfloor \,{=}\, (001)\}$. We represent any such fermionic unit cell graphically as a box containing empty and filled circles, corresponding to the zeros and ones, respectively, in the binary string:
\begin{center}
\begin{tikzpicture}
    \draw[thick] (0,0) rectangle (1,0.5);
    \filldraw[black] (0.25,0.25) circle (2.5pt);
    \filldraw[fill=white] (0.5,0.25) circle (2.5pt);
    \filldraw[fill=white] (0.75,0.25) circle (2.5pt);
    \node[] at (1.2,0) {$,$};
    \draw[thick] (2,0) rectangle (3,0.5);
    \filldraw[fill=white] (2.25,0.25) circle (2.5pt);
    \filldraw[black] (2.5,0.25) circle (2.5pt);
    \filldraw[fill=white] (2.75,0.25) circle (2.5pt);
    \node[] at (3.2,0) {$,$};
    \draw[thick] (4,0) rectangle (5,0.5);
    \filldraw[fill=white] (4.25,0.25) circle (2.5pt);
    \filldraw[fill=white] (4.5,0.25) circle (2.5pt);
    \filldraw[black] (4.75,0.25) circle (2.5pt);
    \node[] at (5.2,0) {$.$};
\end{tikzpicture}
\end{center}
In general, let $k=1,\ldots, K$ label unit cells $\lfloor k \rfloor$ for a given FQH fluid. For an $N$-particle system, we consider $m$ consecutive unit cells, labeled by $q=1,\ldots, m$. For a given $q$, we consider the Hilbert space $\mathcal{H}_q \,{=}\,\mathrm{span}\{|1\rangle, \ldots, |K\rangle\}$, that is, the one with the basis being the moduli space of unit cells, $|k\rangle \,{\equiv}\, |\lfloor k\rfloor\rangle$. To distinguish between different Hilbert spaces, we introduce an additional index $q$ for the states, $|k\rangle_q$ with $k \,{=}\, 1,\ldots, K$. For a given $q$, the domain walls at $q$ are defined as elements of the form $|k\rangle_q \,{\otimes}\, |l\rangle_{q+1} \,{\in}\, \mathcal{H}_q \,{\otimes}\, \mathcal{H}_{q+1}$, for $k,l \,{=}\, 1,\ldots, K$, and denoted by $|k\,l\rangle_q$. If not causing confusion, we will often omit the index $q$. For example, for the $\nu \,{=}\, 1/3$ Laughlin fluid, the set of domain walls (at a fixed $q$) contains 
\begin{eqnarray}\hspace*{-0.5cm}
|1\, 2\rangle = |1\rangle \otimes | 2\rangle\,, \quad |2\, 3\rangle=| 2\rangle\otimes | 3\rangle\,, \quad \cdots . 
\end{eqnarray}

Having defined a set of domain walls, we would like to determine how to fuse them; that is, we are looking for fusion rules associated with a given FQH fluid. For that purpose, we make use of the generalized Schrieffer counting argument to determine first which domain walls are physically independent.

Our approach relies on the observation that, for the Laughlin droplet, an elementary QH located at $z \,{=}\, 0$ is created by acting on the ground state with the operator $e_N$ \cite{Bochniak22}, the second-quantized version of the $N$th elementary symmetric polynomial. In terms of unit cells in the thin-cylinder/torus limit, this reduces to the action of $e_1$. 

We begin by defining the elementary flux-insertion operators acting on the moduli space, with any resulting components outside it discarded:
\begin{eqnarray}
    \mathcal{A}_d^s=e^{\;}_d \, e_{n_{\sf p}}^s &,&  \quad  d \,{=}\, 0,\ldots, n_{\sf p}{-}1, \ s \,{=}\, 0,\ldots, M{-}1 ,
     \nonumber\\
    \mbox{ where } e_d&=&\frac{1}{d!} \!\! \sum_{j_1,\dots ,j_d\ge0}^{M-1} \!\!\!\!
a_{j_1+1}^\dagger  \dots a_{j_d+1}^\dagger a^{\;}_{j_d}\dots a^{\;}_{j_1} ,
\end{eqnarray}
and $e_0 \,{=}\, \mathds{1}$. We assume next that \(n_{\sf p} M\geq K\) \footnote{Otherwise, we need to include non-elementary Schur operators as explained in \cite{supp} in the context of Hafnian fluids. Elementary operators correspond to column representations of symmetric polynomials.}.  Here, $a_j^\dagger$ ($a_j^{\;}$) denotes (properly normalized) fermionic or bosonic creation (anni\-hi\-la\-tion) operators at angular momentum $j$ \cite{Mazaheri2015,Bochniak22}. We impose periodic boundary conditions by identifying $a_M^\dagger \,{\equiv}\, a_0^\dagger$. The operator $e_{n_{\sf p}}$ acts as a global shift, while $e_d$ with $d \,{<}\, n_{\sf p}$ distributes $d$ fluxes in all possible ways among the $n_{\sf p}$ particles. Due to symmetries of the moduli space, it may occur that for some $s_{\max}\,{<}\, M{-}1$ and $d_{\max} \,{<}\, n_{\sf p}{-}1$, the operator $\mathcal{A}_{d_{\max}}^{s_{\max}}$ already acts as $\mathcal{A}_0^0$, so that not all choices need to be considered. We define the class $[s;d]$ as the set $\{(\I \otimes \mathcal{A}_d^s)|k\,k\rangle \,{:}\, \lfloor k \rfloor \,{\in}\,\mathcal{M}\}$. These elementary operators satisfy the particle-hole-type relation $(\mathcal{A}_d^s)^\dagger \sim \mathcal{A}_{n_{\sf p}-d}^{M-s-1}$.
 
We then impose the charge su\-per\-selection rule. We employ Schrieffer's counting argument to associate a charge with each domain wall, that is, we look for a deficiency or excess of physical charge within each unit-cell–length segment of the given domain wall. To illustrate this procedure, consider the domain wall $|(100)\rangle \,{\otimes}\, |(010)\rangle$ in the $\nu \,{=}\, 1/3$ Laughlin fluid, represented graphically as
\begin{center} 
\begin{tikzpicture}
  \draw[thick] (0,0) rectangle (1,0.5);
    \filldraw[black] (0.25,0.25) circle (2.5pt);
    \filldraw[fill=white] (0.5,0.25) circle (2.5pt);
    \filldraw[fill=white] (0.75,0.25) circle (2.5pt);
    \draw[thick] (1,0) rectangle (2,0.5);
    \filldraw[fill=white] (1.25,0.25) circle (2.5pt);
    \filldraw[black] (1.5,0.25) circle (2.5pt);
    \filldraw[fill=white] (1.75,0.25) circle (2.5pt);
    \node[] at (2.2,0) {$.$};
    \draw[thick, dashed, red] (0.375,-0.1) rectangle (1.375,0.6);
\end{tikzpicture}
\end{center}
One readily observes the presence of a size-$3$ box, indicated by the dashed lines, that contains a deficit of filled circles. We must therefore examine every set of three consecutive substrings in the binary string $\ldots 100010 \ldots$. They are $100$, $000$, $001$ and $010$. Since only one of these substrings contains one fewer $1$ than any unit cell in the ground state, we associate this domain wall with a single negative elementary charge, $-e^\ast$, where $e^\ast \,{=}\, e/3$ with $e\,{<}\, 0$. Algebraically, this prescription corresponds to the following charge operator (with eigenvalue $Q \, e^*$):
\begin{equation}
\begin{split}
    \widehat{Q}&=e^\ast \sum\limits_{r=0}^M \Bigl(\sum\limits_{j=r}^{r+M-1}\!n_j -n_{\sf p}\Bigr)\\
    &=e^\ast \Bigl(\sum\limits_{j=0}^{2M-1} \min\{j{+}1,2M{-}j\}\, n_j - n_{\sf p}\,(M\,{+}\,{1})\Bigr) \,,
\end{split}
\end{equation}
$n_j \,{=}\, a_j^\dagger a_j^{\;}$, which is equivalent to the change in macroscopic polarization \cite{mpol1,mpol2}. This is consistent with the observation that fundamental QHs correspond to domain walls carrying elementary charge \cite{Ardonne_2008,Ardonne09}; in other words, topological defects can be identified with generalized domain walls. This leads us to impose the following superselection rule: {\it Domain walls of differing charge belong to distinct classes.} We label the resulting classes by $[s;d;Q]$.

To illustrate this construction, let us consider the $\nu \,{=}\, 1/M$charge of an elementary excitation Laughlin state, where $M  \,{\in}\, 2\mathbb{N}{+}1$. The corresponding moduli space $\mathcal{M}_{\mathrm{L}}^M$ consists of the strings $(10\ldots 0 )$, $(01\ldots 0)$, $\ldots$ and $(0\ldots 01)$, each of which contains exactly one $1$ and $M{-}1$ zeros. A domain wall $|k\,l\rangle$ (with $k,l \,{=}\, 1,\ldots, M$) belongs to one of $2M{-}1$ distinct charge superselection sectors, labeled by $Q \,{=}\, k - l \,{\in}\, \{0,\pm 1,\ldots, \pm(M{-}1)\}$. We immediately observe that, in the neutral sector, only the trivial class $|\I)\!=\!\{|k\,k\rangle\,|\, k \,{=}\, 1,\ldots, M\}$ is present, whereas each of the other charge sectors contains ($Q>0$)
\begin{equation}
\begin{split}
    &|\overline{Q})\!=\!\{|1\,Q{+}1\rangle, |2\,Q{+}2\rangle,\ldots, |M{-}Q\, M\rangle\} \,,\\
    &|{Q})\!=\!\{|Q{+}1\,1\rangle, |Q{+}2\, 2\rangle,\ldots, |M\, M{-}Q\rangle\} \,,
\end{split}
\end{equation}
where $\overline{Q} \,{=}\,{-}Q$. For $M=3$, the charge superselection sectors are summarized in Table~\ref{tab:domain_walls_charge_2}. 
\begin{table}[htb!]
    \centering
    \begin{tabular}{|c|c|}
        \hline 
        {\bf Charge sector} &{\bf Classes }\\ 
         \hline \hline
         neutral & $|\I)=\{|1\,1\rangle,|2\,2\rangle,|3\,3\rangle\}$\\
%         \hline
         $1 \QH$ & $|\overline{1})=\{|1\,2\rangle,|2\,3\rangle\}$ \\
%         \hline
         $2 \QH$ & $|\overline{2})=\{|1\,3\rangle\}$\\
%         \hline
         $2 \QE$ & $|2)=\{|3\,1\rangle\}$\\
%         \hline
         $1 \QE$ & $|1)=\{|2\,1\rangle,|3\,2\rangle\}$\\
         \hline
    \end{tabular}
         \caption{The classes within charge superselection sectors for domain walls in the $\nu \,{=}\, 1/3$ Laughlin fluid.}
    \label{tab:domain_walls_charge_2}
\end{table}

{\it Fusion of domain walls.--}
After incorporating all relevant identifications, we proceed to define the fusion of (the classes of) domain walls. Consider two topological domain walls, $|k_1\,l_1\rangle$  and $|k_2\,l_2\rangle$. Fusion is allowed provided that the respective classes contain representatives such that $l_1 \,{=}\, k_2$. The resulting fused domain wall is $|k_1\,l_2\rangle$. Depending on the particular fluid, this condition may be satisfied in several inequivalent ways, with the resulting domain walls belonging to different classes. Consequently, multiple fusion channels may arise.

Take a $\nu \!=\!1/3$ Laughlin fluid, and suppose one wants to fuse a representative domain wall from class $|2)$,  $|3\,1\rangle \,{=}\, |(001)\rangle \,{\otimes}\, |(100)\rangle$,  with another from class $|\overline{1})$, $|2\,3\rangle \,{=}\, |(010)\rangle \,{\otimes}\, |(001)\rangle$. Since within class $|\overline{1})$ there exists $|1\,2\rangle=|(100)\rangle\otimes |(010)\rangle$, which shares the same vacuum $\lfloor 1 \rfloor$ with $| 3\, 1\rangle$, we fuse $|(001)\rangle \,{\otimes}\, |(100)\rangle$ with $|(100)\rangle \,{\otimes}\, |(010)\rangle$. The result is $|3\,2\rangle=|(001)\rangle \,{\otimes}\, |(010)\rangle$. Graphically, 
\begin{center}
    \begin{tikzpicture}
    \draw[thick] (0,0) rectangle (1,0.5);
    \filldraw[fill=white] (0.25,0.25) circle (2.5pt);
    \filldraw[fill=white] (0.5,0.25) circle (2.5pt);
    \filldraw[fill=black] (0.75,0.25) circle (2.5pt);
    \draw[thick] (1,0) rectangle (2,0.5);
    \filldraw[fill=black] (1.25,0.25) circle (2.5pt);
    \filldraw[fill=white] (1.5,0.25) circle (2.5pt);
    \filldraw[fill=white] (1.75,0.25) circle (2.5pt);
    \node[] at (2.5,0.25) {$\times$};  
    \draw[thick] (3,0) rectangle (4,0.5);
    \filldraw[fill=black] (3.25,0.25) circle (2.5pt);
    \filldraw[fill=white] (3.5,0.25) circle (2.5pt);
    \filldraw[fill=white] (3.75,0.25) circle (2.5pt);
    \draw[thick] (4,0) rectangle (5,0.5);
    \filldraw[fill=white] (4.25,0.25) circle (2.5pt);
    \filldraw[fill=black] (4.5,0.25) circle (2.5pt);
    \filldraw[fill=white] (4.75,0.25) circle (2.5pt);
    \node[] at (5.5,0.25) {$=$};
    \draw[thick] (6,0) rectangle (7,0.5);
    \filldraw[fill=white] (6.25,0.25) circle (2.5pt);
    \filldraw[fill=white] (6.5,0.25) circle (2.5pt);
    \filldraw[fill=black] (6.75,0.25) circle (2.5pt);
    \draw[thick] (7,0) rectangle (8,0.5);
    \filldraw[fill=white] (7.25,0.25) circle (2.5pt);
    \filldraw[fill=black] (7.5,0.25) circle (2.5pt);
    \filldraw[fill=white] (7.75,0.25) circle (2.5pt);
    \node[] at (8.25,0.25) {$.$};
\end{tikzpicture}
\end{center}
In Table~\ref{tab:fusionZ3}, we summarize all possible fusions of classes that we obtain when proceeding along these lines.

\begin{table}[htb!]
    \centering
    \begin{tabular}{| c |  c |}
        \hline 
        {\bf Fused types} &{\bf Fusion channels}\\ 
         \hline \hline 
         $|\overline{1}) \times |\overline{1}) $ & $|\overline{2})$\\
%         \hline
         $|\overline{1}) \times |2)$ & $|1)$\\
%         \hline
         $|\overline{1})\times |1)$ & $|\I)$\\
%         \hline
         $|\overline{2})\times |2)$ & $|\I)$\\
%         \hline
         $|\overline{2})\times |1)$ & $|\overline{1})$\\
%         \hline
         $|1)\times |1)$ & $|2)$\\
         \hline
    \end{tabular}
         \caption{Fusion rules in the $\nu \,{=}\, 1/3$ Laughlin fluid. The remaining, trivial, fusion rules involve the neutral anyon $|\I)$.}
    \label{tab:fusionZ3}
\end{table}

Some classes do not admit fusion; for example, one can\-not fuse $|\overline{1})$ with $|\overline{2})$, or $|{1})$ with $|{2})$. These commutative rules are independent of the choice of representative, and imply that, whenever defined, only a single fusion channel is present. We thus obtain a model of Abelian anyons.

The apparent obstruction to fusing certain classes is resolved by noting that charge modularity, $\mathrm e^{\mathrm i 2\pi Q}$, must be preserved. In terms of fusion categories, imposing this requirement amounts to considering modules over a suitable algebra in the category (see the End Matter). In the present framework, this is realized by identifying domain walls that differ by a local fermion or boson, i.e., those which are related by the action of the operator $\mathcal{A}_0^s \,{\otimes}\, \mathcal{A}_0^s$ for some $s \,{=}\, 0, \ldots, M{-}1$. For the $\nu \,{=}\, 1/3$ Laughlin fluid, this leads to a reduction of the previously obtained classes into the following ones: $|\I)_\star \,{=}\, \{|1\,1\rangle, |2\,2\rangle, |3\,3\rangle\}$, $|\overline{1})_\star \,{=}\, \{|1\,2\rangle, |2\,3\rangle, |3\,1\rangle\}$ and $|\overline{2})_\star \,{=}\, \{|2\,1\rangle, |3\,2\rangle, |1\,3\rangle\}$.

Our next task is to identify the mathematical structure underlying these fusion rules. At first sight, one might anticipate a relation with the group $\mathbb{Z}_3 \,{=}\, \{\mathbb{I}, a, a^2\}$. However, inspired by CFT considerations, the presence of five distinct charge sectors suggests that this structure should be extended by $\mathbb{Z}_5 \,{=}\, \{-2, -1, 0, 1, 2\}$. This naturally leads to the product group $\mathbb{Z}_3 \,{\times}\, \mathbb{Z}_5 \,{=}\, \{a^\mu_q\}^{\mu=0,1,2}_{q=-2,-1,\ldots,2}$, with the corresponding fusion rules given by
\begin{equation}
    a^\mu_{q_1} \times a^\nu_{q_2} = a^{[\mu+\nu]_3}_{[q_1+q_2]_5},
\end{equation}
with $[\ell]_p$ denoting $\ell\,\pmod{p}$. However, clearly, not all elements of the product group correspond to physical sectors. The allowed elements are
\begin{equation*}
\begin{split}
    &a^0_0 = |\mathds{1}), \quad 
    a^1_{-1} = |\overline{1}), \quad 
    a^1_{2} = |2), \quad
    a^2_{1} = |1), \quad 
    a^2_{-2} = |\overline{2}).
\end{split}
\end{equation*}
By imposing charge modularity we identify $a^1_{-1} \,{=}\, a^1_{2}$ and $a^2_{1} \,{=}\, a^2_{-2}$, which obeys the $\mathbb{Z}_3$ fusion rules \footnote{The above construction readily generalizes to arbitrary $M$. In this case, one starts from $\mathbb{Z}_M \,{\times}\, \mathbb{Z}_{2M{-}1}$, with fusion rules $a^\mu_{q_1} \times a^\nu_{q_2} = a^{[\mu+\nu]_M}_{[q_1+q_2]_{2M-1}}$,
which, after identifications and charge modularity, leads to $\mathbb{Z}_M$ fusion rules. 
}.

{\it Abelian fluids.--} Another physically relevant class of FQH states arises within the framework of composite fermions (CFs) \cite{Jain_book}. These states are known to accurately describe FQH fluids with filling fraction $\nu \,{=}\, n/(2np+1)$, where $n$ and $p$ are integers, leading to the so-called Jain sequence. The LLL projected wave function describing these states is given by ($\mathcal{P}_{\rm LLL}$ stands for the projector)
\begin{equation}
\label{eq:wf2_5}
    \Psi_{\!\frac{n}{2np+1}}(Z_N)=\mathcal{P}_{\rm LLL} \prod\limits_{i<j} (z_i-z_j)^{2p} \Psi_{n}(Z_N) \,,
\end{equation}
where $Z_N \,{=}\, \{z_1,\ldots, z_N\}$ denotes the set of particle coordinates, defined on the plane, and $\Psi_{n}(Z_N)$ is the $N$-par\-ticle wave function of the integer quantum Hall state at filling fraction $n$, involving $n$ filled LLs. The single-par\-ticle Landau orbitals, in the $n$th LL, read 
\begin{equation}
    \eta_{n,j}(z^\ast,z)\!=\!\frac{(-1)^n\sqrt{n!}\, z^j}{\sqrt{2\pi 2^j (n+j)!}}\, L_n^j\biggl(\frac{|z|^2}{2}\biggr)\, \mathrm e^{-\frac{|z|^2}{4}},
\end{equation}
where $L_n^j$ are the generalized Laguerre polynomials \cite{Jain_book}. 

%\cite{ChenYang20}
We begin with the $\nu \,{=}\, 2/5$ case, corresponding to $n \,{=}\, 2$ and $p \,{=}\, 1$.  Applying the algorithm described above to the $N \,{=}\, 4$ particle wave function, we obtain the ground state root pattern $1010010100$. Within the CF framework~\cite{Jain_book}, an elementary QH is created by removing one electron from the highest occupied orbital \footnote{One may ask what happens if, instead of following the CF prescription for generating QHs, one removes an electron from the LLL rather than from the highest occupied orbital. These two distinct constructions lead to markedly different physical consequences: while the CF-based construction gives rise to Abelian anyons, the alternative construction does not.}. At the level of the wave function, this amounts to replacing $\Psi_{2}(Z_4)$ in Eq.~\eqref{eq:wf2_5} with the QH wave function $\Psi_{2}^{\mathrm{qh}}(Z_4)$, given by the determinant
\begin{equation}
\begin{vmatrix}
        \eta_{0,0}(z_1)& \eta_{0,0}(z_2) & \eta_{0,0}(z_3) & \eta_{0,0}(z_4)\\
        \eta_{0,1}(z_1) & \eta_{0,1}(z_2) & \eta_{0,1}(z_3) & \eta_{0,1}(z_4)\\
        \eta_{1,1}(z_1,z_1^\ast) & \eta_{1,1}(z_2,z_2^\ast) & \eta_{1,1}(z_3,z_3^\ast) &
        \eta_{1,1}(z_4,z_4^\ast)\\
        \eta_{1,2}(z_1,z_1^\ast) & \eta_{1,2}(z_2,z_2^\ast) & \eta_{1,2}(z_3,z_3^\ast) &
        \eta_{1,2}(z_4,z_4^\ast)
    \end{vmatrix}.
\end{equation}
The corresponding root pattern is given by $1001010010$. Proceeding in this manner and accounting for all QH configurations, we conclude that the moduli space $\mathcal{M}_{\mathrm{Jain}}^{2/5}$ consists of $\lfloor 1\rfloor \,{=}\, (10100)$, $\lfloor 2\rfloor \,{=}\, (01010)$, $\lfloor 3\rfloor \,{=}\, (00101)$, $\lfloor 4 \rfloor \,{=}\, (10010)$, and $\lfloor 5 \rfloor \,{=}\, (01001)$. 

Using the flux-insertion operators defined above, one can define the classes $[s; d; Q]$. A straightforward calculation~\cite{supp} shows that there are nine such classes. The fusion rules among these classes take the form ($Q_1, Q_2 \,{>}\, 0$)
\begin{equation}
    \begin{split}
        |\overline{Q}_1) \times |\overline{Q}_2)&=|\overline{Q_1+Q_2}), \quad 1\leq Q_1+Q_2\leq 4,\\
        |Q_1) \times |Q_2)&=|Q_1+Q_2), \quad 1\leq Q_1+Q_2\leq 4,\\
        |\overline{Q}) \times |Q )&=|\I), \hspace*{2cm}  1\leq Q \leq 4,\\
        |\overline{Q}_1)\times |Q_2)&=|\overline{Q_1-Q_2}), \quad 1\leq Q_2<Q_1\leq 4,\\
        |\overline{Q}_1)\times |Q_2)&=|Q_2-Q_1),\quad 1\leq Q_1<Q_2\leq 4.
    \end{split}
\end{equation}

Again, not all fusions are permitted; for instance, the fusion between $|\overline{2})$ and $|\overline{3})$ is not possible. The fusion structure coincides with that of the $\nu \,{=}\, 1/5$ Laughlin fluid, yielding a subset of $\mathbb{Z}_5 \,{\times}\, \mathbb{Z}_9$. Adopting the same notation as in the Laughlin case, we identify $a^0_0 \,{=}\, \mathds{1}$ and $a^Q_{-Q} \,{=}\, |\overline{Q})$, $a^Q_{5-Q} \,{=}\, |Q)$, with $Q \,{=}\, 1, \ldots, 4$. Imposing charge modularity (equivalence under fusion with a local fermion) reduces these to the classes \cite{supp} that satisfy the familiar $\mathbb{Z}_5$ fusion rules. An analogous analysis for the $\nu \,{=}\,2/9$ and $\nu \,{=}\,3/7$ Jain states is provided in \cite{supp}.

More generally, if all elements $\lfloor k\rfloor \,{\in}\, \cal M$ are related by translations, i.e., $|[k{+}s]_M \rangle \,{=}\, {\cal A}^s_0\, |k\rangle$, then the fusion of all domain wall classes is Abelian \footnote{ For those cases the possible charges are $Q_s(k)=M\sum_{j=0}^{s-1} n^k_{M-j} - s \, n_{\sf p}$, where $\lfloor k \rfloor =(n^k_1,n^k_2,\ldots,n^k_M)$ and $s=0,1,\ldots,M-1$.}. Note that in this case, $|k \, [k{+}s]_M \rangle \,{\times}\,  |l \, [l{+}\tilde s]_M \rangle \,{=}\, |k \, [k{+}s{+}\tilde s]_M \rangle$ if $l \,{=}\, [k{+}s]_M$, and since there is a unique class for each translation, there is at most one fusion outcome. 

{\it Non-Abelian fluids.--}
In all Abelian cases analyzed above, the moduli space is generated by a single element and its translations. The simplest example of a fluid expected to host non-Abelian anyonic excitations is the $\nu \,{=}\, 1/2$ Pfaffian system, also known as the Moore--Read state~\cite{Moore91}. We now derive its fusion rules within our microscopic framework. We find that the corresponding moduli space $\mathcal{M}_{\mathrm{Pf}}$ consists of six elements: $\mathcal{M}_{\mathrm{Pf}}$: $\lfloor 1\rfloor \,{=}\, (0101)$, $\lfloor 2\rfloor \,{=}\, (1010)$, $\lfloor 3\rfloor \,{=}\, (1100)$, $\lfloor 4\rfloor \,{=}\, (0110)$, $\lfloor 5\rfloor \,{=}\, (0011)$, and $\lfloor 6\rfloor \,{=}\, (1001)$. In contrast to the Abe\-lian examples,  $\mathcal{M}_{\mathrm{Pf}}$ contains elements not connected by the action of translations.  Applying the flux-insertion operators $\mathcal{A}_d^s$ and using Schrieffer-like charge counting, we obtain twelve classes $[s; d; Q]$ (see \cite{supp}). Among these, there are two distinct classes with $Q \,{=}\, 0$: the trivial vacuum $|\I)$ and $|\I)_0 \,{=}\, \{\ket{1\,1}, \ket{2\,2}, \ket{4\,6}, \ket{6\,4}\}$. We also find two classes with charge $Q \,{=}\, {-}2$,
\begin{align*}
    |\overline{2})_1 & =[1;0;-2]=\{\ket{2\,1},\ket{3\,4},\ket{4\,5}\}\,,\\
    |\overline{2})_2 &= [3;0;-2] = \{\ket{2\,1},\ket{3\,6},\ket{6\,5}\}\,,
\end{align*}
and similarly for $Q=2$. We then derive the fusion rules between these domain walls -- see Table~I in \cite{supp} for details. In particular, the fusion of two elementary QHs (each with charge -$e/4$) has two channels:
\begin{equation}
     |\overline{1})\times |\overline{1})=  |\overline{2})_1\oplus |\overline{2})_2\,,
\end{equation}
indicating the non-Abelian nature of these excitations. Taking charge modularity into account, we obtain six distinct classes. This classification does not alter the fact that two distinct classes of two QHs remain present. 

We next compare our results with those obtained from CFT, where the system is described by a $\mathbb{Z}_2$ Ising parafermion model enriched by an electric charge~\cite{Cappelli10}. By selecting appropriate topologically distinct operators, one identifies, for given pairs of operators $b$ and $c$, all operators $a$ satisfying fusion relations of the form $a \,{\times}\, b \,{=}\, c \oplus \ldots$. This construction yields twelve distinct classes labeled by $a$, which can be mapped onto those derived from our microscopic analysis. Further details are provided in~\cite{supp}. In this way, we establish a direct connection to the Ising fusion rules, long conjectured to govern excitations in the Pfaffian fluid. The CFT description is also closely related to the framework developed in~\cite{Nayak11}, as discussed in~\cite{supp}.

For the remaining examples, whenever a candidate CFT description is available, we benchmark our microscopic results against it. The Pfaffian fluid can be viewed as a particular member of the Read--Rezayi (RR) family of states~\cite{Read99}. Consider the $\nu \,{=}\, 3/5$ RR state, which can be obtained by antisymmetrizing \cite{Bochniak23} Halperin’s $333111$ state~\cite{Halperin83}. The moduli space $\mathcal{M}_{\mathrm{RR}}^{3/5}$ contains ten elements: $\mathcal{M}_{\mathrm{RR}}^{3/5}$: $\lfloor 1\rfloor \,{=}\,(10110)$, $\lfloor 2\rfloor \,{=}\,(01011)$, $\lfloor 3\rfloor \,{=}\, (10101)$, $\lfloor 4\rfloor =  (11010)$, $\lfloor 5\rfloor \,{=}\, (01101)$, $\lfloor 6\rfloor \,{=}\, (11100)$, $\lfloor 7\rfloor \,{=}\, (01110)$, $\lfloor 8\rfloor \,{=}\, (00111)$, $\lfloor 9\rfloor \,{=}\, (10011)$ and $\lfloor 10\rfloor \,{=}\, (11001)$. Our procedure yields $22$ classes of domain walls $[s; d; Q]$ and their associated fusion rules. Imposing charge modularity reduces the number of distinct classes to ten. Once again, we find perfect agreement of our microscopic derivation with the CFT benchmark based on charged parafermions. Further details are provided in~\cite{supp}.

{\it Topological fluids of bosons.--}
The fluids discussed thus far correspond to fermionic systems. However, the procedure extends straightforwardly to bosons. We illustrate this in the case of the Gaffnian fluid \cite{Simon07}. The corresponding moduli space $\mathcal{M}_{\mathrm{Gaff}}$ consists of six unit cells: $\lfloor 1 \rfloor=(200)$, $\lfloor 2\rfloor = (020)$, $\lfloor 3 \rfloor = (002)$, $\lfloor 4 \rfloor = (110)$, $\lfloor 5\rfloor \,{=}\, (011)$ and $\lfloor 6 \rfloor \,{=}\, (101)$. Analyzing the action of the flux-insertion operators yields fourteen classes, which we summarize —together with their fusion rules— in~\cite{supp}. Among these, there are two classes with $Q \,{=}\, 0$, two with $Q \,{=}\,{\pm} 1$ and $Q \,{=}\, {\pm} 2$, and single classes with charges $Q \,{=}\, {\pm} 3$ and $Q \,{=}\, {\pm} 4$. Taking charge modularity into account reduces the number of distinct classes to six.

The CFT underlying this fluid is the $\mathcal{W}_2(3,5)$ minimal model, enriched by the free boson CFT $U(1)_6$. This minimal model is known to be isomorphic to the Virasoro minimal model with central charge $c \,{=}\, {-}3/5$, which is non-unitary. By computing the operator product expansions for the six primary fields in the product model, one derives the fusion rules for the corresponding anyons. In this framework, the six anyons correspond to the six ground states of the Gaffnian fluid. For any two ground states $b$ and $c$, one can identify a corresponding $a$ satisfying $a \,{\times}\, b \,{=}\, c \,{\oplus}\, \ldots$. This procedure defines $14$ classes, which are in one-to-one correspondence with those derived from microscopic principles.  

{\it The Hafnian: A new frontier.--}
Finally, we apply our method to the Hafnian, the densest zero-energy state of a local three-body Hamiltonian, for which no CFT or TQFT description is known. 

The bosonic Hafnian at $\nu \,{=}\, 1/2$ is believed to possess an infinite number of distinct quasiparticle sectors (and is accordingly deemed to be an ``irrational theory'' \cite{Hermanns2011}), a feature reflected in its torus ground-state degeneracy, which grows with the number of particles, $N$. 
Using the thin-torus root configurations of Ref.~\cite{Hermanns2011} as input, we examine the anyon fusion algebra implied by our method. The resulting moduli space naturally decomposes into two sectors. The first, consisting of all configurations satisfying the ${\leq}2$-in-$4$ rule, coincides with the root configurations of the $(2,4)$ Jack polynomial \cite{Bernevig08}. Restricted to this sector, our construction closes and produces the fusion algebra of $\mathfrak{su}(4)_2$. In this case, however, the moduli space contains $K=10$ states, while $n_{\sf p}M=8$, requiring two additional flux insertion operators beyond the elementary $\mathcal{A}^s_d$ \cite{supp}. This perfectly matches the fact that the $\mathfrak{su}(4)_2$ fusion is obtained when starting from the product fusion rules
$\mathcal Z_{8}^{(1/2)} \,{\boxtimes}\, \mathfrak{su}(2)_4$
and extending by the commutative algebra
$0 \,{\boxtimes}\, 0 \,{\oplus}\, 4 \,{\boxtimes}\, 4$. The second sector is generated by the squeezed-Laughlin defect cell $(02001)$, which violates the ${\leq}2$-in-$4$ rule and whose multiplicity \emph{appears to grow} with $N$ \cite{Hermanns2011}. Once these defect configurations are included, our construction no longer closes to a finite fusion ring, regardless of system size. 

The fermionic Hafnian at $\nu \,{=}\, 1/3$ is equivalent to the antisymmetrized Halperin's $551$ state~\cite{Jeong2017,Bochniak23}. In this case, $K=15$ and  $n_{\sf p}M=12$, again requiring (three) additional flux-insertion operators \cite{supp}. The algebra generated by our flux-insertion operators closes. The resulting fusion ring is obtained by extending $\mathcal Z_{12}^{(1/2)} \,{\boxtimes}\, \mathfrak{su}(2)_4$ by a supercommutative algebra, analogous to the fermionic Pfaffian case (End Matter).

{\it Conclusions and Outlook.--} 
We developed an algebraic framework that yields the fusion rules of anyonic quasiparticles in FQH systems directly from microscopic information encoded in the DNA of candidate many-body ground-state wave functions. By leveraging the structure of dominant orbital occupation patterns and generalizing Schrieffer’s counting argument via appropriate superselection relations among excitations, our approach yields a unified derivation of fusion rules for both Abelian and non-Abelian fluids \footnote{Our algebraic framework applies broadly to fluids beyond the examples considered here, including the full fermionic and bosonic Read–Rezayi sequence. }. It thereby provides a direct connection between microscopic input and emergent topological order in fermionic and bosonic systems alike. When available, our framework reproduces constructions within CFT and category theory (see the End Matter and the Supplemental Material). More broadly, our results establish a systematic route for obtaining topological data from first principles and open the door to the analysis of more general strongly correlated quantum phases.

Looking ahead, our framework suggests several promising directions for further investigation. A key avenue is the systematic reconstruction of the full braided fusion category data directly from microscopic input, thereby extending the present derivation of fusion rules to encompass braiding statistics and modular properties. Establishing such a connection would provide a fully microscopic route to the categorical formulation of topological order. Another important direction is to move beyond holomorphic wave functions,  enabling the treatment of more general quantum Hall states in which Landau-level mixing plays a significant role.

Our ultimate goal is a comprehensive microscopic theory of topological phases and defects, linking wave-func\-tion-based constructions with the full algebraic and categorical framework of topological quantum matter.

{\it Acknowledgements.--} We thank J.~Jain and N.~Snyder for useful comments.  The work of A.B. was partially funded by the Deutsche Forschungsgemeinschaft (DFG, Ger\-man Research Foundation) under Germany’s Excellence Strategy - EXC-2111 – 390814868.  A.B. is also supported by the Alexander von Humboldt Foundation. G.O. gratefully acknowledges support from the Institute for Advanced Study. J.F. is supported by VR under project no.~2022-02931. S.R. is supported by a Simons Investigator Grant from the Simons Foundation (Award No. 566116). This work is supported by the Gordon and Betty Moore 
Foundation EPiQS initiative, Grant GBMF8685.01. This work was performed in part at Aspen Center for Physics, which is supported by National Science Foundation grant PHY-2210452.

\bibliography{apssamp}

\section*{End Matter} 

Here, we provide a TQFT perspective on pertinent features of the microscopic analysis performed in the main text. This perspective also helps to clarify the interpretations and physical pictures that underlie the microscopic scheme. To avoid confusion between the microscopic and TQFT formulations, we reserve the terms ``topological excitation'', ``domain wall'' and ``topological ground state'' for the microscopic context. By contrast, when we use the term ``anyon”, we are using the TQFT description or, in other words, refer to a simple object in the corresponding category. 

In the main text, we introduce a framework for creating domain walls of DNAs through the rules based on the flux-insertion mechanism. These rules not only generate domain walls, but they also organize them into classes. Once the classes are identified, we determine how they can be fused. From the TQFT perspective, the flux-insertion rules can be interpreted as arising from bulk anyons that act as (generically non-invertible) symmetries on the ground states.

A suitable mathematical structure for describing this situation consists of a fusion category $\mathcal C$ and a module category $\mathcal M$ over $\mathcal C$. For a mathematical background on this structure, we refer to  \cite{Etingof-book}. In our context, $\mathcal C$ is the category of non-invertible symmetries; we denote its simple objects by $a, b, c, ...$\,. We will often refer to these objects also as symmetry lines or as line operators. While our discussion primarily takes place in the $(1{+}1)$d limit -- the thin-cylinder/torus or Tao–Thouless limit -- of $(2{+}1)$d systems, in the full $(2{+}1)$d picture these lines correspond to bulk anyons. The simple objects of the module category $\mathcal M$ represent DNAs or, equivalently, ground states of the system; we denote them by $ k, l, m, ...$\,. They typically constitute gapped ground states of a suitable $(1{+}1)$d parent Hamiltonian.

We note that module categories arise naturally in various branches of physics, like, e.g., Levin-Wen models with boundary \cite{Kitaev_Kong_2012}, MPO symmetries in string-net models \cite{Lootens21}, or conformal field theories on surfaces with boundary \cite{Fuchs2002}. In the latter context, $\mathcal C$ is the category of topological defect lines, while $\mathcal M$ is the category of maximally symmetric boundary conditions.

That $\mathcal C$ is a fusion category means, in particular, that the line operators obey fusion rules
  \begin{equation}
  a \otimes b \cong \bigoplus_c N_{a,b}^{c}\, c \,,
  \label{eq:a.b=...}
  \end{equation}
with non-negative integer multiplicities $N_{a,b}^{c}$. Likewise, that $\mathcal M$ is a module category over $\mathcal C$ means that the line operators can act on the ground states. The action of a line operator $a$ on the ground state $k$ decomposes as
  \begin{equation}
  a \triangleright k \cong \bigoplus_l M_{a,k}^{l}\, l \, ,
  \label{eq:a.k=...}
  \end{equation}
with non-negative integers $M_{a,k}^{l}$. This relation parallels our microscopic rules, which, when acting on a ground state yield a collection of ground states as output. Put differently, the microscopic rules provide a concrete way to compute the mixed fusion coefficients $M_{a,k}^{l}$. We express the fusion rules \eqref{eq:a.b=...} and \eqref{eq:a.k=...} diagrammatically as

\begin{equation*}
    \begin{tikzpicture}[
  midarrow/.style={
    postaction={
      decorate,
      decoration={
        markings,
        mark=at position 0.5 with {\arrow{Stealth}}
      }
    }
  }
]

\draw[midarrow] (0,0) -- (1,0);
\draw[midarrow] (0,0.5) -- (1,0.5);

\node at (0.5,-0.25) {$b$};
\node at (0.5,0.75) {$a$};

\node at (1.85,0.15) {$=\bigoplus\limits_{c} N^c_{a,b}$};

\draw[midarrow] (2.75,0.25) -- (3.75,0.25);

\node at (3.25,0.5) {$c$};
\node at (3.9,0.2) {$,$};

\draw[midarrow] (4.25,0) -- (5.25,0);
\draw[midarrow] (4.25,0.5) -- (5.25,0.5);

\node at (4.75,-0.25) {$k$};
\node at (4.75,0.75) {$a$};

\node at (6.15,0.15) {$=\bigoplus\limits_{c} M^l_{a,k}$};

\draw[midarrow] (7,0.25) -- (8,0.25);

\node at (7.5,0.5) {$l$};
\node at (8.2,0.2) {$,$};
\end{tikzpicture}
\end{equation*}
respectively. Here, the stacking of the two lines indicates the fusion of two line operators and the action of a line operator on a ground state, respectively.

An alternative notation for the action $\triangleright$ that is sometimes used is the tensor product symbol $\otimes$, but this notation is strictly appropriate only when the module category $\mathcal M$ under consideration is the regular $\mathcal C$-module category, namely \cite{Etingof-book} $\mathcal C$ regarded as a module over itself, meaning that $a$ and $k$ are to be treated as labels of the same type. This is the case, for example, when the non-invertible symmetry $\mathcal C$ is completely broken by the ground states. In the cases considered in the main text, we are indeed dealing with the regular module category. Nevertheless, it is worthwhile to keep this distinction in mind -- one may suspect that there will be natural generalizations in which other module categories arise.

When the action of $a$ on $k$ generates $l$, i.e.,  if we have $a \,{\triangleright}\, k \,{\cong}\, l \,{\oplus}\, ...\,$, then the line operator $a$ can end at the domain wall between $k$ and $l$, giving rise to a topological junction,
\begin{equation*}
    \begin{tikzpicture}[
  midarrow/.style={
    postaction={
      decorate,
      decoration={
        markings,
        mark=at position 0.5 with {\arrow{Stealth}}
      }
    }
  }
]
 \draw[midarrow] (0,0) -- (1,0);   \draw[midarrow] (1,0) -- (2,0);
 \draw[midarrow] (1,1) -- (1,0);

 \node at (0.5,-0.25) {$k$};
 \node at (1.5,-0.25) {$l$};
 \node at (0.8, 0.75) {$a$};
 \node at (2.1,0) {$.$};   
    \end{tikzpicture}
\end{equation*}
In other words, we deal with a topological operator at the domain wall; the vector space of such operators is (isomorphic to) the space $\mathrm{Hom}_{\mathcal M}(a \,{\triangleright}\, k,l)$ of module morphisms. Note that if the integer $M_{a,k}^{l} = \dim(\mathrm{Hom}_{\mathcal M}(a \,{\triangleright}\, k,l))$ is larger than $1$, then we should have more than one rule leading to the domain wall $|k\,l\rangle$, as observed in the main text for several examples.

Let us now explain how, given the mixed fusion coefficients $M_{a,k}^{l}$, which can be calculated from the microscopic rules, we can obtain the ordinary fusion rules $N_{a,b}^{c}$. In the particular case of the regular module category $\mathcal M \,{=}\, \mathcal C$, we simply have $N_{a,b}^{c} \,{=}\, M_{a,b}^{c}$. In the general case, we use the fact that, by the definition of a module category, if the output of $b \,{\triangleright}\, k$ and the input of $a \,{\triangleright}\, l$ match, then we can combine or fuse $b$ and $a$:

\begin{equation*}
    \begin{tikzpicture}[
  midarrow/.style={
    postaction={
      decorate,
      decoration={
        markings,
        mark=at position 0.5 with {\arrow{Stealth}}
      }
    }
  }
]
 \draw[midarrow] (0,0) -- (1,0);   \draw[midarrow] (1,0) -- (2,0);
 \draw[midarrow] (1,1) -- (1,0);

 \node at (0.5,-0.25) {$k$};
 \node at (1.5,-0.25) {$l$};
 \node at (0.8, 0.75) {$b$};
 \node at (2.5,0.35) {$\times$};

 \draw[midarrow] (3,0) -- (4,0);   \draw[midarrow] (4,0) -- (5,0);
 \draw[midarrow] (4,1) -- (4,0);

 \node at (3.5,-0.25) {$l$};
 \node at (4.5,-0.25) {$m$};
 \node at (3.8, 0.75) {$a$};
 \node at (5.5,0.35) {$=$};

 \draw[midarrow] (6,0) -- (7,0);   \draw[midarrow] (7,0) -- (8,0);
 \draw[midarrow] (7,1) -- (7,0);

 \node at (6.5,-0.25) {$k$};
 \node at (7.5,-0.25) {$m$};
 \node at (6.4, 0.75) {$b\otimes a$};
 \node at (8.1, 0) {$.$};
    \end{tikzpicture}
\end{equation*}
In formulas, this means that the expressions
  \begin{equation}
  a \,{\triangleright}\, (b \,{\triangleright}\, k)
  = a \,{\triangleright}\, \Big( \sum_l M_{b,k}^{l}\, l \Big)
  = \sum_{l,l'} M_{b,k}^{l}\, M_{a,l}^{l'}\, l'
  \end{equation}
and
  \begin{equation}
  (a \,{\otimes}\, b) \,{\triangleright}\, k
  = \sum_c N_{a,b}^{c}\, c \,{\triangleright}\, k
  = \sum_{c,l} N_{a,b}^{c}\, M_{c,k}^{l}\, l 
  \end{equation}
coincide, i.e.,\ we have the equality
  \begin{equation}
  \sum_{c} N_{a,b}^{c}\, M_{c,k}^{l'} = \sum_{l} M_{b,k}^{l}\, M_{a,l}^{l'}
  \label{NM=MM}
  \end{equation}
for all objects $a,b \,{\in}\, \mathcal C$ and $k,l' \,{\in}\, \mathcal M$. It is convenient to combine the information about all domain walls that can be created by using the microscopic rule corresponding to $a$ into a matrix $M_a$, having entries $(M_a)_k^l \,{=}\, M_{a,k}^{l}$. Then the equality \eqref{NM=MM} is recast into the matrix equation
  \begin{equation}
   \sum_c N_{a,b}^{c}\, M_c^{} = M_a^{}\, M_b^{} \,. 
  \end{equation}
Given the numbers $M_{a,k}^{l}$, we can solve this equation to obtain the $N_{a,b}^{c}$. 

{\it The (fermionic) Pfaffian fluid.}-- Here we present a categorical interpretation of the results obtained in the main text for the Pfaffian fluid. The starting point is the category  $  \Ufour \boti \Is $, where $\Is$ is the Ising category and $\Ufour$ captures the electric charge. $\Is$ has three sectors (i.e., isomorphism classes of simple objects), which we denote by $\{I, \sigma, \psi\}$, with quantum dimensions
  \begin{equation}
  d_I = 1 = d_\psi \,, \quad d_\sigma = \sqrt2 
  \end{equation}
and statistical phases
  \begin{equation}
  \theta_1 = 1 \,, \quad \theta_\psi = -1 \,, \quad
  \theta_\sigma = \mathrm e^{\pi\mathrm i/8} .
  \end{equation}
The non-trivial fusion rules are
  \begin{equation}
  \psi \oti \psi \cong I \,, \quad \psi \oti \sigma \cong \sigma \,, \quad
  \sigma \oti \sigma \cong I \oplus \psi \,.
  \label{eq:furuIs}
  \end{equation}
The category $\Ufour$ has eight sectors, which we denote by $V_q$ with
$q \,{\in}\, \{-3,-2,...\,,4\}$. Each of them has unit quan\-tum dimension, their electric charges are $q/4$, and their statistical phases are $ \theta_{V_q} \,{=}\, \mathrm e^{\pi\mathrm iq^2/8}$.
$\Ufour$ has $\mathbb Z_8$ fusion rules, $ V_q \oti V_{q'} \,{\cong}\, V_{q+q'} \,$, with the labels counted modulo $8$ \footnote{These $\Ufour$ data are, e.g., realized by the $\mathfrak u(1)_8$ conformal field theory.}. The electron corresponds to the sector $V_4 \boti \psi$. To account for its {\it triviality}, we identify it with the tensor unit $V_0 \boti I$. Concretely, we
introduce the object
  \begin{equation}
  A := V_0 \boti I \oplus V_4 \boti \psi \,,
  \end{equation}
which has a natural structure of a supercommutative algebra in the category $ \Ufour \boti \Is $, and consider a specific class of $A$-modules. The algebra $A$ has $12$ inequivalent simple modules. We admit only the $6$ modules for which $q$ is even when $V_q$ is combined with $I$ or $\psi$, while $q$ is odd when $V_q$ is combined with $\sigma$. One of these admitted modules is $A$ itself; the others have the form   
\begin{equation}
  \begin{aligned}
  & V_3 \boti \sigma \oplus V_{-1} \boti \sigma \,, \quad
  V_1 \boti \sigma \oplus V_{-3} \boti \sigma \,, 
  \\[3pt]
  & V_2 \boti I \oplus V_{-2} \boti \psi \,, \quad
  V_2 \boti \psi \oplus V_{-2} \boti I \,,
  \\[3pt]
  & V_0 \boti \psi \oplus V_4 \boti I \,. 
  \end{aligned}
  \label{eq:nonlocAmodules}
  \end{equation}
The \emph{ground states} correspond to those simple subobjects of these modules that in the conformal field theory realization of the category $ \Ufour \boti \Is$ have the lower conformal weight. These are 
  \begin{align}
  & \lfloor 1 \rfloor \equiv V_{-1}\boti \sigma\,,\quad \lfloor 2\rfloor \equiv V_1\boti \sigma\,, \quad \lfloor 3\rfloor \equiv V_2\boti I,\notag\\[2pt]
  & \lfloor 4 \rfloor \equiv V_0\boti I\,, \quad \lfloor 5\rfloor \equiv V_{-2}\boti I\,, \quad \lfloor 6\rfloor \equiv V_0\boti \psi\,.
  \label{eq:def:VI...}
  \end{align}
The resulting fusion rules $M_{a,k}^{l}$ follow directly from the ones in $\Is$ and $\Ufour$, accounting also for the module structure in \eqref{eq:nonlocAmodules}. We are in the situation that the $M$- and $N$-matrices coincide. The matrix $M_{\lfloor 4\rfloor}$ is the $6\times 6$ unit matrix; further (when numbering rows and
columns according to \eqref{eq:def:VI...}),
\begin{align}
    &M_{\lfloor 6 \rfloor} =\begin{pmatrix}
        1 & 0 & 0 & 0 & 0 & 0\\
        0 & 1 & 0 & 0 & 0 & 0\\
        0 & 0 & 0 & 0 & 1 & 0\\
        0 & 0 & 0 & 0 & 0 & 1\\
        0 & 0 & 1 & 0 & 0 & 0\\
        0 & 0 & 0 & 1 & 0 & 0
    \end{pmatrix}, \quad M_{\lfloor 3 \rfloor} = \begin{pmatrix}
        0 & 1 & 0 & 0 & 0 & 0\\
        1 & 0 & 0 & 0 & 0 & 0\\
        0 & 0 & 0 & 0 & 0 & 1\\
        0 & 0 & 1 & 0 & 0 & 0\\
        0 & 0 & 0 & 1 & 0 & 0\\
        0 & 0 & 0 & 0 & 1 & 0
    \end{pmatrix},\notag\\
    & M_{\lfloor 2 \rfloor} = \begin{pmatrix}
        0 & 0 & 0 & 1 & 0 & 1\\
        0 & 0 & 1 & 0 & 1 & 0\\
        1 & 0 & 0 & 0 & 0 & 0\\
        0 & 1 & 0 & 0 & 0 & 0\\
        1 & 0 & 0 & 0 & 0 & 0\\
        0 & 1 & 0 & 0 & 0 & 0
    \end{pmatrix}, \quad 
    \begin{array}{l}
    M_{\lfloor 5 \rfloor}=M_{\lfloor 3 \rfloor}^T\,, \\~\\ M_{\lfloor 1\rfloor} = M_{\lfloor 2\rfloor}^T \,. \end{array}
\end{align}
%%%%%%%%%%%%%%%%%%%%%%%%%%%%%%%%%

In a rational CFT, the modular S-matrix diagonalizes the fusion rules.
In general, this property does not determine the S-matrix uniquely, but it does so (up to a permutation of the columns that is an automorphism of the fusion rules) when imposing further properties, namely symmetry, unitarity, and $S_{0,a}\,{\ge}\,0$. In the present case the result is (cf. Example 2.16 of Ref. \cite{Bruillard2017})
  \begin{equation} \hspace{-0.2cm}
  S = \frac12 \, \begin{pmatrix}
  \sqi &    1      & \sqi &    \sqi   &    1      &    \sqi   \\[3pt]
     1 &    0      &   -1 & \mathrm i &    0      &-\mathrm i \\[3pt]
  \sqi &   -1      & \sqi &    \sqi   &   -1      &    \sqi   \\[3pt]
  \sqi & \mathrm i & \sqi &   -\sqi   &-\mathrm i &   -\sqi   \\[3pt]
     1 &    0      &   -1 &-\mathrm i &    0      & \mathrm i \\[3pt]
  \sqi &-\mathrm i & \sqi &   -\sqi   & \mathrm i &   -\sqi   \end{pmatrix} ,
  \end{equation}
where the ordering of rows and columns corresponds to order the objects in Eq.\ \eqref{eq:def:VI...} (and analogously for the modules in \eqref{eq:nonlocAmodules}) as
$\lfloor 4 \rfloor$, $\lfloor 2 \rfloor$, $\lfloor 6 \rfloor$,
$\lfloor 5 \rfloor$, $\lfloor 1 \rfloor$, $\lfloor 3 \rfloor$.

%%%%%%%%%%%%%%%%%%%%%%%%%%%%%%%%%

\end{document}